\documentclass{JHEP3}
\usepackage{graphicx}

\usepackage{array}

\usepackage{amsmath}
\usepackage{amsfonts}
\usepackage{amssymb}

\usepackage{feynmp}

\renewcommand{\=}[1][=]{\mathrel{\phantom{#1}}}

\DeclareMathOperator{\tr}{tr}
\newcommand{\del}{\partial}
\newcommand{\diff}[2][\,]{\mathrm{d}^{#1\!}#2}
\newcommand{\pint}[2][d]{\int\!\!\frac{\diff[#1]{#2}}{(2\pi)^#1}}

\newcommand{\kaD}[1][]{{\ka_{(D)}^{#1}}}
\newcommand{\emD}[1][]{{M_{(D)}^{#1}}}

\newcommand{\emd}[1][]{{M_{(d)}^{#1}}}

\renewcommand{\L}{\mathcal{L}}
\renewcommand{\O}{\mathcal{O}}

\newcommand{\hh}{\hat{h}}
\newcommand{\hs}{\tilde{h}}

\newcommand{\nmode}[1][n]{{(\vec{#1})}}
\newcommand{\Nnmode}[1][n]{{(-\vec{#1})}}

\newcommand{\al}{\alpha}
\newcommand{\be}{\beta}
\newcommand{\ga}{\gamma}
\newcommand{\de}{\delta}

\newcommand{\ka}{\kappa}
\newcommand{\si}{\sigma}

\newcommand{\Ga}{\Gamma}
\newcommand{\La}{\Lambda}

\title{Gravitational Contributions to the Running Yang-Mills Coupling in Large Extra-Dimensional Brane Worlds}
\author{Dietmar Ebert$^{a,b}$, Jan Plefka$^{b}$ and Andreas Rodigast$^{b}$ \\ 
$^a$Joint Institute for Nuclear Research, Dubna, R-141980, Russia\\ 
$^b$Humboldt-Universit\"at zu Berlin, Institut f\"ur Physik, Newtonstra{\ss}e 15, D-12489 Berlin, Germany\\ 
E-mail: \href{mailto:debert@physik.hu-berlin.de}{\tt debert},\href{mailto:plefka@physik.hu-berlin.de}{\tt plefka},\email{rodigast@physik.hu-berlin.de}}
\preprint{HU-EP-08/32}

\abstract{We study the question of a modification of the running gauge coupling of Yang-Mills theories
due to quantum gravitational effects in a compact large extra dimensional brane world scenario with a low energy 
quantum gravity scale. The ADD scenario is applied for a $D=d+\delta$ dimensional space-time in which gravitons
freely propagate, whereas the non-abelian gauge fields are confined to a $d$-dimensional brane. The extra dimensions
are taken to be toroidal and the transverse fluctuation modes (branons) of the brane are taken 
into account. On this basis we have calculated the one-loop corrections due to  virtual Kaluza-Klein graviton 
and branon modes for the gluon two- and three-point functions in an effective field theory treatment.
Applying momentum cut-off regularization we find that for a $d=4$ brane the leading 
gravitational divergencies cancel irrespective of the number of 
extra dimensions $\delta$, generalizing previous results in the absence of extra-dimensions. 
Hence, again the Yang-Mills $\beta$-function receives \emph{no} gravitational corrections at one-loop.
This is no longer true in a `universal' extra dimensional scenario with a $d>4$ dimensional brane.
Moreover, the subleading  power-law gravitational divergencies induce higher-dimensional counterterms, 
which we establish in our scheme. Interestingly, for $d=4$ these gravitationally 
induced counterterms are of the form recently considered 
in non-abelian Lee-Wick extensions of the standard model -- now with a possible mass scale in the TeV range
due to the presence of large extra dimensions.
}

\begin{document}

\begin{fmffile}{graphs}

\section{Introduction}

Large extra dimensions with a low energy scale for quantum gravity \cite{ADD} 
represent a much discussed resolution of the hierarchy problem between the Planck scale 
$M_\text{Planck}\sim 10^{19}\text{GeV}$ and the electroweak scale $M_\text{weak}\sim 1 \text{TeV}$ 
of the standard model of particle physics\footnote{For a set of recent reviews see \cite{Reviews}}. 
Present experimental constraints allow for $\delta\geq 2$ 
extra dimensions of up to submillimeter size if one insists on a fundamental gravitational scale of
a few TeV being just in the range of the upcoming LHC collider at CERN \cite{PDG}. 
In this scenario the standard model fields (or their supersymmetric extensions) are confined to 
a 4 dimensional brane within a $D=4+\delta$ dimensional
space-time manifold with compact extra dimensions in which the gravitons freely propagate. Extra dimensions 
arise in superstring theories and such braneworld scenarios can be embedded within string theory
\cite{AADD}. However, 
they may also be studied within quantum field theory upon treating gravity as an effective field theory. 
As the inclusion of gravity
to the standard model (or its supersymmetric extensions) destroys its renormalizability in four
dimensions, one might just as well also consider the existence of extra ``universal'' compact dimensions for
the brane fields. This was proposed for the first time in \cite{Antoniadis:1990ew}. 
This universal extra dimensions scenario in the absence of gravity was considered 
by the authors of \cite{Dienes} 
who showed that the presence of extra dimensions for the minimal supersymmetric standard model (MSSM) fields 
leads (with a suitable cut-off procedure for the Kaluza-Klein towers of states)
to a power law running of the MSSM couplings and grand unification at scales $M\ll 10^{16} 
\text{GeV}$ well below the standard grand unified scale. A natural question to be addressed in this
work is then how the running of gauge couplings is affected once one includes quantum gravitational effects
in such a brane-world scenario.
 
Recently, the question of gravitational contributions to the running gauge couplings in \emph{four} dimensional
Einstein-Yang-Mills theory has received considerable attention. This was initiated by the work of Robinson and 
Wilczek \cite{RobWil} who reported a one-loop contribution to the Yang-Mills $\beta$-function from 
virtual gravitons yielding a dominant power-law running behaviour for any gauge theory at energies close 
to the Planck scale. However, this was later on shown by Pietrykowski and Toms \cite{Pietrykowski,Toms} to be a 
gauge artefact of the background field method employed in \cite{RobWil}. 
A reanalysis using background field techniques \cite{Pietrykowski,Toms} as well as an unambigous diagramatic 
approach employing a momentum cut-off regularization to be sensitive to non-logarithmic divergencies
of the present authors \cite{EPR} demonstrated the \emph{absence} of gravitational
contributions to the Yang-Mills $\beta$-function in four dimensions.\footnote{This vanishing was shown to also occur
in a string theoretical analysis for certain 4d $\mathcal{N}=1$ supersymmetric compactifications using an infrared regulator
\cite{Elias}.} 
Despite this, the Einstein-Yang-Mills theory receveives counterterm corrections of dimension six
\cite{Deser,EPR} arising from the, by power counting subleading,
logarithmic divergencies due to virtual gravitons at one-loop.  
Interestingly, the induced non-abelian gluonic counterterm is of Lee-Wick \cite{LeeWick} form 
$M^{-2}_{\text{Planck}}\tr[D^\mu F_{\mu\rho}D_\nu F^{\nu\rho}]$, which has recently 
been independently considered as a non-abelian 
Lee-Wick extension of the standard model in order to stabilize the Higgs mass against qudratically divergent
radiative corrections \cite{Grinstein}\footnote{This phenomenon in the \emph{abelian} case was also noted
in \cite{WuZhong}.}. However, while being of conceptual interest this effect
is tiny at TeV scales due to the largeness of the Planck mass in the absence of extra dimensions.

Motivated by the large extra dimensional scenario of Arkani-Hamed, Dimopoulos and Dvali (ADD) \cite{ADD} we have
extended our earlier four dimensional investigation \cite{EPR} 
to the most general scenario of a $D=d+\delta$ dimensional brane-world
with a $d$ dimensional Yang-Mills brane theory embedded in a $D$ dimensional manifold in which the graviton
propagates. For simplicity the extra dimensions are taken in the form of a $\delta$-torus $T^\delta$ 
with common radii $R$.
Viewed from the brane we have a tower of Kaluza-Klein graviton excitations contributing in the considered
one-loop effective theory. Following \cite{ContRatt} we also take into account brane fluctuations.
Due to the invariance under general coordinate transformations, the theory then contains Goldstone 
bosons (branons) which interact with the Kaluza-Klein states and have to be included into the investigations.
We perform a recalculation of the gluon two- and three-point functions at one-loop in the extra dimensional
setup and determine the gravitational contributions to the
Yang-Mills $\beta$-function from the leading divergencies of the momentum 
cut-off regulated integrals. 
Interestingly it is shown that \emph{again} the leading bulk gravitational corrections cancel for a $d=4$ dimensional
Yang-Mills brane theory. Moreover we establish the necessary counterterms for the subleading divergencies,
which could be viewed as gravity induced Lee-Wick extensions of the theory. Here we point out subtle ambiguities in the
calculation of subleading power-like divergencies and offer a (universal) resolution of these ambiguities by invoking
gauge invariance.
 In this framework and for a $D=4+\delta$ dimensional brane-world
the non-abelian counterterm remains of the non-abelian Lee-Wick form 
$M^{-2}_{(D)}\, \tr[D^\mu F_{\mu\rho}D_\nu F^{\nu\rho}]$, where $M_{(D)}$ is the low gravitational scale
of a few TeV. One is then naturally tempted to attribute the non-abelian Lee-Wick extension considered
by Grinstein O'Conell and Wise \cite{Grinstein} with this brane-world quantum gravity counterterm. However,
this is to be taken with caution as higher-loop gravitational 
corrections will introduce an infinite tower of higher-dimensional counterterms, 
thus modifying the result of \cite{Grinstein}.

It should be noted that the results for the dimensionful coefficents of the higher derivative terms and the dimensionful gauge coupling on $d\!\neq\!4$ dimensional branes and their renormalization depend on the choice of the graviton gauge condition \cite{ContRatt,Antoniadis:1985ub}. Therefore the obtained values for these cases are specific to the applied de~Donder gauge. Nevertheless, all actual observables, like scattering amplitudes, should be independent of the chosen gauge as shown in \cite{Antoniadis:1985ub}.

\section{General Formalism}\label{sc:formalism}
\subsection{Effective Lagrangians for the Einstein-Yang-Mills theory and branes}

Let us consider gravity in $D$-dimensional space-time $\mathcal{M}=\mathbb{R}^{1,d-1}\times T^{\de}$,
where $T^{\de}$ is a ${\de}$-dimensional torus with a uniform radius $R$
and $\operatorname{dim}\mathcal{M}=D=d+\de$. We decompose the metric as
\begin{equation}\label{eq:Metric_decomp}
 G_{M N}=\eta_{M N}+\kaD h_{M N}
\end{equation}
around the flat $D$-dimensional Minkowski space-time with
$\eta_{MN}=\operatorname{diag}(+,-,\dots,-)$ in terms of the graviton field $h_{M N}$. Here
$\kaD[2]=32\pi/\emD[D-2]$ is the gravitational coupling constant in $D$-space-time dimension with
$\emD$ being the corresponding low scale Planck mass. This low scale Planck mass is related to 
the Planck mass observed on the brane $\emd=M_\text{Planck}\sim 10^{19}\text{GeV}$ as
\[
 \emD[D-2](2\pi R)^\de =M_\text{Planck}^{d-2} .
\]
 In the following upper (lower) case latin letters are used for $D$-dimensional
($\de$-dimensional compactified) indices and Greek letters for $d$-dimensional indices. 
Let us further decompose the $D$-dimensional coordinates as
$X^M=(x^\mu,z^i)$ and write the field $h_{M N}$ in matrix form
\begin{equation}
 h_{M N}=\begin{pmatrix}
      \hh_{\mu\nu}-\tfrac{1}{d-2}\phi\,\eta_{\mu\nu} & \tfrac{1}{\sqrt{2}}B_{i\,\nu} \\
           \tfrac{1}{\sqrt{2}}B_{j\,\mu}         &      \phi_{ij}
     \end{pmatrix},
\end{equation}
where we have introduced the fields which appear in the $d$-dimensional effective theory,
 i.~e. the graviton~$\hh_{\mu\nu}$,
graviphotons~$B_{i\,\mu}$ and graviscalars~$\phi_{ij}$ and further have used
$\phi = \eta^{ij}\phi_{ij}=-\de_{ij}\phi_{ij}$, $\mu = 0,1,...,d-1$, $i = d,...,{\de}$.

Next, consider the bulk action of gravity
\begin{equation}
 S=\int\diff[D]{X} (\frac{2}{\ka_{(D)}^2} \sqrt{-G} \,\mathcal{R} -\frac{1}{\alpha}F_N F^N + \L_\text{ghosts}) . \\
\end{equation}
Here $F_N$ denotes the gauge fixing term
\[
 F_N=\del^\mu\left(h_{\mu N}-\tfrac{1}{2}\eta_{\mu N}h\right)+\alpha\del^i\left(h_{i N}-\tfrac{1}{2\alpha}\eta_{i N}h\right)
\]
with $h=h^M_M$ and $\al$ being the gauge parameter. In particular, we will consider here the de~Donder gauge
$\al=1$ which is more suitable for our loop calculations than the unitary gauge, $\al\to\infty$. The propagation of unphysical fields in the de~Donder gauge does not pose problems since we consider no processes with external gravitational fields and the effective number of degrees of freedom is the same in all gauges. On the other hand in de~Donder gauge the propagators yield a better UV-behaviour than in unitary gauge. 
Note that the gravitational Faddeev-Popov ghosts will play no role in the considered one-loop calculations, so there is no need to specify $\L_\text{ghost}$.

Decomposing now the bulk action around the flat background and taking into account only the quadratic part in
the gravitational field leads to the quadratic bulk Lagrangian \cite{GiuRattWells,HanLykkZh}
\begin{equation}\label{eq:grav-lag}
 \L^{(D)}_\text{grav.}=\tfrac{1}{2}\del_A h_{M N} \left(\eta^{M R}\eta^{N S}-\tfrac{1}{2}\eta^{M N}\eta^{R S}\right)\del^A
 h_{R S} \,,
\end{equation}
where $\eta_{MN}$ is used to raise and lower indices. It is convenient to perform the Kaluza-Klein reduction of this
Lagrangian by decomposing the field $h_{MN}(x,z)$ which is compactified on the
${\de}$-dimensional torus $T^{\de}$ into the mode expansion
\begin{equation}\label{eq:modes}
 h_{MN}(x,z) = V_{\delta}^{-1/2} \sum_{\vec{n}\in\mathbb{Z}^\delta} h_{MN}^{(\vec{n})}(x) e^{i\frac{\vec{n}\cdot\vec{z}}{R}} ,
\end{equation}
where $V_{\delta}=(2\pi R)^{\delta}$ is the volume of the compactified torus.

By integrating the Lagrangian \eqref{eq:grav-lag} over the compactified extra coordinates, one obtains the $d$-dimensional
Lagrangian for Kaluza-Klein graviton states. The quadratic part of this Lagrangian reads:
\begin{equation}\label{eq:lag_grav_exp}
\begin{aligned}
 \L^{(d)}_\text{grav.}=\frac{1}{2}\sum_{\vec{n}}\Bigl(& \del_\al \hh^{\nmode}_{\mu\nu} \left(\eta^{\mu\rho}\eta^{\nu\si}-
								\tfrac{1}{2}\eta^{\mu\nu}\eta^{\rho\si}\right) \del^\al \hh^{\Nnmode}_{\rho\si} - \\
       & -m^2_{\vec{n}} \hh^{\nmode}_{\mu\nu} \left(\eta^{\mu\rho}\eta^{\nu\si}-\tfrac{1}{2}\eta^{\mu\nu}\eta^{\rho\si}\right)\hh^{\Nnmode}_{\rho\si} - \\
       & -\del_\al B^{\nmode}_{i\,\mu} \del^\al B_{i}^{\Nnmode\,\mu} + \\
       & +m^2_{\vec{n}} B^{\nmode}_{i\,\mu} B_{i}^{\Nnmode\,\mu} + \\
       & +\del_\al \phi^{\nmode}_{ij} \left(\de_{ik}\de_{jl}+\tfrac{1}{d-2}\de_{ij}\de_{kl}\right)\del^\al \phi^{\Nnmode}_{kl} - \\
       & -m^2_{\vec{n}} \phi^{\nmode}_{ij} \left(\de_{ik}\de_{jl}+\tfrac{1}{d-2}\de_{ij}\de_{kl}\right) \phi^{\Nnmode}_{kl} \Bigr) ,
\end{aligned}
\end{equation}
where $m^2_{\vec{n}}=\vec{n}\!\cdot\!\vec{n}/R^2$ is the mass squared of the $\rm{n^{th}}$ excited Kaluza-Klein graviton. From this we can read off the Feynman rules for the propagators of the gravitational fields in de~Donder gauge:
\begin{equation}
\begin{array}{rcl@{\: :\quad}l}
 \hh_{\al\be}^{\nmode}&\begin{fmfgraph}(30,10)
                        \fmfleft{i}\fmfright{o}\fmf{dbl_wiggly}{i,o}
                       \end{fmfgraph}
&\hh_{\ga\de}^{(\vec{n}')} & \dfrac{i\de_{\vec{n},-\vec{n}'}\frac{1}{2}\left(\eta_{\al\ga}\eta_{\be\de}+\eta_{\al\de}\eta_{\be\ga} -\tfrac{2}{d-2}\eta_{\al\be}\eta_{\ga\de}\right)}{p^2-m^2_{\vec{n}}}\\
   B_{i\,\mu}^{\nmode}&\begin{fmfgraph}(30,10)
                        \fmfleft{i}\fmfright{o}\fmf{dbl_curly}{i,o}
                       \end{fmfgraph}
&  B_{j\,\nu}^{(\vec{n}')} & \dfrac{-i\de_{\vec{n},-\vec{n}'}\de_{ij}\eta_{\mu\nu}}{p^2-m^2_{\vec{n}}}\\
    \phi_{ij}^{\nmode}&\begin{fmfgraph}(30,10)
                        \fmfleft{i}\fmfright{o}\fmf{dbl_plain}{i,o}
                       \end{fmfgraph}
& \phi_{kl}^{(\vec{\vec{n}}')}   & \dfrac{i\de_{\vec{n},-\vec{n}'}\frac{1}{2}\left(\de_{ik}\de_{jl}+\de_{il}\de_{jk} -\tfrac{2}{D-2}\de_{ij}\de_{kl}\right)}{p^2-m^2_{\vec{n}}}.
\end{array}
\end{equation}

In contrast to the gravitons moving freely in the bulk, the matter fields (we focus here only on gauge bosons) are confined to a $d$-dimensional space-time manifold (a $(d\!-\!1)$-brane).
In particular, we shall use the brane coordinates
\[
 Y^N(x^\mu)=(y^\mu(x)\!=\!x^\mu,\tfrac{1}{\sqrt{\tau}}\xi^i(x)) , 
\]
where as discussed by Sundrum \cite{Sundrum} the reparameterization invariance of the $(d\!-\!1)$-brane allows to fix $d$ of the coordinates and choose a static gauge
$y^\mu\!=\!x^\mu$.
The $\xi^i$ are dynamical \emph{branon} fields representing the transversal fluctuations of the brane 
forming the Goldstone scalars in $d$ dimensions of the broken translation invariance in the $\de$ 
extra dimensions. $\tau$ is the brane tension introduced at this point to yield a canonical normalization 
for the branons (see below)\footnote{We discard here the interesting question of 
how such a brane can dynamically arise as a solution of the
underlying Einstein-Yang-Mills system or a more general supergravity theory related to string theory.
It is worth mentioning that the perturbation in the gravitational dynamics due to the brane tension is 
small in the region of validity of the effective field theory, see 
appendix~\ref{appendix}.}.

Let us next consider the induced metric on the brane
\begin{equation}\label{eq:ind_metric}
\begin{aligned}
 g_{\mu\nu}(x)&= \frac{\del Y^M}{\del x^\mu}\frac{\del Y^N}{\del x^\nu}G_{M N}(Y(x)) \\
          &= G_{\mu\nu}(x,\tfrac{\xi(x)}{\sqrt{\tau}})+\frac{1}{\sqrt{\tau}}\left(\del_\mu\xi^i G_{i\nu}(x,\tfrac{\xi(x)}{\sqrt{\tau}}) +\del_\nu\xi^i G_{\mu i}(x,\tfrac{\xi(x)}{\sqrt{\tau}}) \right) + \\
           &\= +\frac{1}{\tau}\left(\del_\mu\xi^i \del_\nu\xi^j G_{i j}(x,\tfrac{\xi(x)}{\sqrt{\tau}})\right) \\
           &= G_{\mu\nu}(x,0) \\
           &\= +\frac{1}{\sqrt{\tau}}\left(\xi^i\del_i G_{\mu\nu}(x,0)+\del_\mu\xi^i G_{i\nu}(x,0)+\del_\nu\xi^i G_{\mu i}(x,0)\right) + \\
           &\= +\frac{1}{\tau}\bigl(\tfrac{1}{2}\xi^i\xi^j\del_i\del_j G_{\mu\nu}(x,0)+\xi^i\del_\mu\xi^j\del_i G_{j\nu}(x,0) + \\
           &\phantom{=+\frac{1}{\tau}\bigl(}+\xi^i\del_\nu\xi^j \del_i G_{\mu j}(x,0)+\del_\mu\xi^i \del_\nu\xi^j G_{i j}(x,0)\bigr) +
\O(\tau^{-3/2})
\end{aligned}
\end{equation}
and again decompose it around the flat background
\begin{equation}\label{eq:metric_decomp}
 g_{\mu\nu}=\eta_{\mu\nu}+\ka\, \hs_{\mu\nu} ,
\end{equation}
using now the $d$-dimensional gravitational constant $\ka=\ka_{(d)}=\kaD/(2\pi R)^{\de/2}$. The metric fluctuation $\hs$ has to be expressed in terms of the branon $\xi$ and the $D$-dimensional graviton $h_{MN}$. By plugging equations \eqref{eq:Metric_decomp} and \eqref{eq:modes} into \eqref{eq:ind_metric} one obtains
\begin{equation}\label{eq:h_tilde}
\begin{aligned}
 \hs_{\mu\nu} &= -\frac{1}{\ka\tau}\de_{i j}\del_\mu\xi^i \del_\nu\xi^j + \\
        &\=+\sum_{\vec{n}}\Bigl(\hh^{\nmode}_{\mu\nu} -\tfrac{1}{d-2}\eta_{\mu\nu}\phi^{\nmode} 
		+\frac{1}{\sqrt{\tau}}\Big(\tfrac{i}{R}n_i\xi^i \hh^{\nmode}_{\mu\nu} -\tfrac{i}{R(d-2)}\eta_{\mu\nu}n_i\xi^i\phi^{\nmode}+  \\
	&\phantom{=+\sum\Bigl(\hh^{\nmode}_{\mu\nu} -\tfrac{1}{d-2}\eta_{\mu\nu}\phi^{\nmode} +\frac{1}{\sqrt{\tau}}}	
		+\tfrac{1}{\sqrt{2}}\del_\mu\xi^i B^{\nmode}_{i\,\nu}+ \tfrac{1}{\sqrt{2}}\del_\nu\xi^i B^{\nmode}_{i\,\mu}\Bigr)\Bigr) + \ldots\, 
\end{aligned}
\end{equation}
where the colons refer to trems involving more than two fields.

\subsection{Interaction with gauge bosons and branons}
Let us now consider the brane Lagrangian \cite{ContRatt}
\[
 \L_\text{brane}=\sqrt{-g}\left(-\tau + \L_\text{YM} \right) ,
\]
where $\L_\text{YM}$ describes the covariant $d$-dimensional Yang-Mills Lagrangian of the gauge bosons,
\[
 \L_\text{YM}=-\tfrac{1}{2} g^{\mu \rho} g^{\nu\sigma} \tr\left(F_{\mu\nu} F_{\rho\sigma}\right) ,
\]
with $F_{\mu\nu}=\del_\mu A_\nu - \del_\nu A_\mu -ig [A_\mu,A_\nu]$
the gauge field strength, $A_\mu=A^a_\mu T^a$ the gauge field, $T^a$ the Lie algebra generators of the gauge group and $g$ the gauge boson coupling\footnote{Since  $F_{\mu\nu}$  is antisymmetric the Christoffel 
connections arising from the space-time covariant derivatives $\nabla_{\mu}$ cancel against each other, 
hence the covariant derivatives can be replaced here by ordinary derivatives $\partial_\mu$.}

For our purposes it is convenient to expand the first term in $\L_\text{brane}$ up to quadratic order in
graviton and branon fields which yields
\begin{align} 
 -\tau\sqrt{-g} &= -\tau\Bigl(1 + \frac{\ka}{2}\sum_{\vec{n}}\left(\hh^{\nmode}-\tfrac{d}{d-2}\phi^{\nmode}\right) + \notag \\
		&\phantom{=-\tau\Bigl(}+ \frac{\ka^2}{8}\sum_{\vec{n}}\sum_{\vec{m}}\left( \hh^{\nmode}\hh^{\nmode[m]}
			-2\hh^{\nmode\,\al\be}\hh^{\nmode[m]}_{\al\be}-2\hh^{\nmode}\phi^{\nmode[m]}+\tfrac{d}{d-2}\phi^{\nmode}\phi^{\nmode[m]}\right)\Bigr) + \notag \\
         &\= +\frac{1}{2}\de_{i j}\del^\mu\xi^i \del_\mu\xi^j - \label{eq:brane_tension_exp} \\
         &\= -\frac{\ka\sqrt{\tau}}{2}\sum_{\vec{n}}\left(\tfrac{i}{R}n_i\xi^i\hh^{\nmode}-\tfrac{id}{R(d-2)}n_i\xi^i\phi^{\nmode}+\sqrt{2}B^{\nmode}_{i\,\mu}\del^\mu\xi^i\right) +\L_\text{interaction}\,. \notag
\end{align}
Note that equation~\eqref{eq:brane_tension_exp} contains terms linear in $\hh$ and $\phi$, because the massive brane is a source of gravity. These terms reflect the off-shell nature of the metric expansion,  but they can be neglected for 
the regime under consideration, see  Appendix~\ref{appendix}. Furthermore we find a kinetic term for the massless branons, graviton-branon mixing terms as well as interaction terms. The corresponding Feynman rules are
\begin{equation}
\begin{array}{rcl@{\: :\quad}l}
 \xi^{i}&\begin{fmfgraph}(40,10)
                        \fmfleft{i}\fmfright{o}\fmf{dashes}{i,o}
                       \end{fmfgraph}
&       \xi^{j}            & \dfrac{i\de^{ij}}{p^2} \,, \\
 \hh_{\mu\nu}^{\nmode} &\begin{fmfgraph}(40,10)
                        \fmfleft{i}\fmfright{o}\fmf{dbl_wiggly}{i,v}\fmf{dashes}{v,o}\fmfdot{v}
                       \end{fmfgraph}
& \xi^i & \dfrac{\ka\sqrt{\tau}}{2}\eta_{\mu\nu}\dfrac{n_i}{R} \,, \\
  B_{j\,\mu}^{\nmode}  &\begin{fmfgraph}(40,10)
                        \fmfleft{i}\fmfright{o}\fmf{dbl_curly}{i,v}\fmf{dashes}{v,o}\fmfdot{v}
                       \end{fmfgraph}
& \xi^i & \dfrac{\ka\sqrt{\tau}}{\sqrt{2}}\de_i^j p_\mu \,, \\
  \text{and}\quad \phi_{kl}^{\nmode}  &\begin{fmfgraph}(40,10)
                        \fmfleft{i}\fmfright{o}\fmf{dbl_plain}{i,v}\fmf{dashes}{v,o}\fmfdot{v}
                       \end{fmfgraph}
& \xi^i & \dfrac{\ka\sqrt{\tau}}{2}\dfrac{d}{d-2}\de_{kl}\dfrac{n_i}{R} \,,
\end{array}
\end{equation}
where $p_\mu$ is the incoming momentum of the graviphoton~$B_{j\,\mu}$ and $n_i$ the 
$i$'th mode number of the Kaluza-Klein field.

The interaction of gravitons and branons with the gauge bosons is contained in the covariant dependence
of $\L_\text{YM}$ on the induced metric. The Feynman rules can now be obtained by using equations~\eqref{eq:metric_decomp} and \eqref{eq:h_tilde}.

The first order couplings of the branon and gravitational fields to brane fields are mediated by
\[
 \L^{(\ka)}=-\frac{\ka}{2}T^{\mu\nu}\hs_{\mu\nu}
\]
and shown in figure~1. Here $T^{\mu\nu}$ is the energy-momentum tensor of the brane fields.
\begin{figure}[t]
\label{fig:vertices}%
\vspace{3ex}
\begin{center}
\begin{tabular}{*{3}{@{\qquad}m{40\unitlength}@{}>{$\displaystyle}l<{$}}@{\qquad}}
 \begin{fmfgraph*}(40,40)
  \fmfbottom{i,o}\fmf{boson}{i,v,o}\fmftop{t}\fmfv{l=$\hh^{\nmode}_{\al\be}$}{t}\fmf{dbl_wiggly}{v,t}\fmfdot{v}
 \end{fmfgraph*}&=-i\frac{\ka}{2}T_{\al\be} &
 \begin{fmfgraph*}(40,40)
  \fmfbottom{i,o}\fmf{boson}{i,v,o}\fmftop{t}\fmfv{l=$\phi^{\nmode}_{ij}$}{t}\fmf{dbl_plain}{v,t}\fmfdot{v}
 \end{fmfgraph*}&=-i\frac{\ka}{2}\frac{1}{d-2}T_{\mu}^{\mu}\de_{ij} &
 \begin{fmfgraph*}(40,40)
  \fmfbottom{i,o}\fmf{boson}{i,v,o}\fmftop{t}\fmfv{l=$B^{\nmode}_{i\al}$}{t}\fmf{dbl_curly}{v,t}\fmfdot{v}
 \end{fmfgraph*}&= 0 \\[7ex]
 \begin{fmfgraph*}(40,40)
  \fmfbottom{i,o}\fmf{boson}{i,v,o}\fmftop{t1,t2}\fmfv{l=$\xi^i$}{t1}\fmfv{l=$\xi^j$}{t2}\fmf{dashes}{t1,v,t2}\fmfdot{v}
 \end{fmfgraph*}&=-i\frac{1}{\tau}T_{\mu\nu}\de_{ij}k^\mu_1 k^\nu_2 &
 \begin{fmfgraph*}(40,40)
  \fmfbottom{i,o}\fmf{boson}{i,v,o}\fmftop{t1,t2}\fmfv{l=$\xi^i$}{t1}\fmfv{l=$\hh^{\nmode}_{\al\be}$}{t2}\fmf{dashes}{t1,v}\fmf{dbl_wiggly}{v,t2}\fmfdot{v}
 \end{fmfgraph*}&=\frac{\ka}{2\sqrt{\tau}}T_{\al\be}\frac{n_i}{R} \\[7ex]
 \begin{fmfgraph*}(40,40)
  \fmfbottom{i,o}\fmf{boson}{i,v,o}\fmftop{t1,t2}\fmfv{l=$\xi^i$}{t1}\fmfv{l=$\phi^{\nmode}_{kl}$}{t2}\fmf{dashes}{t1,v}\fmf{dbl_plain}{v,t2}\fmfdot{v}
 \end{fmfgraph*}&=\frac{\ka}{2\sqrt{\tau}}\frac{1}{d-2}T_{\mu}^{\mu}\de_{kl}\frac{n_i}{R} &
 \begin{fmfgraph*}(40,40)
  \fmfbottom{i,o}\fmf{boson}{i,v,o}\fmftop{t1,t2}\fmfv{l=$\xi^i$}{t1}\fmfv{l=$B^{\nmode}_{j\al}$}{t2}\fmf{dashes}{t1,v}\fmf{dbl_curly}{v,t2}\fmfdot{v}
 \end{fmfgraph*}&=-\frac{\ka}{\sqrt{2 \tau}}T_{\mu\al}\de_i^j k^\mu 
\end{tabular}
\end{center}
\caption{Vertices of graviton--matter and branon--matter couplings; $k_1$, $k_2$, $k$ are the incoming momenta of the branons (gluons are drawn as example).}
\end{figure}

The higher order couplings of the Kalzua-Klein gravitons $\hh^{\nmode}$ and graviscalars  $\phi^{\nmode}_{ij}$ to the brane fields are derived from the couplings of the well studied $d$~dimensional graviton $h_{\mu\nu}$ which couples identically as the composed field $\hs$ in the higher dimensional case. From equation~\eqref{eq:h_tilde} one finds that the Kaluza-Klein gravtions $\hh$ can be substituted for $d$~dimensional $h$'s. For each $d$~dimensional graviton $h_{\mu\nu}$ which is substituted by a graviscalar $\phi^{\nmode}_{ij}$ the corresponding vertex rule is obtained by multiplying the orginal formula by $\de_{ij}\eta^{\mu\nu}/(d-2)$. The explicit rules can be taken e.\,g. from \cite{EPR}.

\section{Results}\label{sc:results}
\subsection[Gravitational contributions to the $\beta$-function]{\boldmath Gravitational contributions to the $\beta$-function}
In this chapter we will calculate the leading divergencies of the gauge boson (gluon) propagator
arising from one-loop diagrams with exchange of virtual Kaluza-Klein gravitons ($\hh^{(\vec{n})}$-- spin 2,
$B_{j\mu}^{(\vec{n})}$-- spin 1,$\phi^{(\vec{n})}$-- spin 0) and branons ($\xi^i$)
(To be precise, note that the spin-1 KK-graviton does not couple to gauge bosons).
Although we are only interested in pure gravitational one-loop contributions and will not consider any terms involving the brane tension $\tau$, we cannot ignore the branons completely due to the inverse $\tau$ dependence in matter--branon interactions and graviton--branon mixing \cite{ContRatt}, see figure~1 and equation~\eqref{eq:brane_tension_exp}. There exist two shapes of tadpole graphs involving branons at order $\ka^2$. These were calculated using the general form of the interaction, figure~1, so the results are applicable for generic brane matter fields. 
In particular, we get 
\fmfstraight
\begin{equation}\label{eq:parachute}
\begin{aligned}
 \begin{fmfgraph}(30,30)
 \fmfbottom{i,v,o}\fmfleft{v1}\fmfright{v2}\fmf{dashes}{v,v1}\fmf{dbl_wiggly,left}{v1,v2}\fmf{dashes}{v2,v}\fmf{boson}{i,v,o}\fmfdot{v1,v2,v}
\end{fmfgraph}+
\begin{fmfgraph}(30,30)
 \fmfbottom{i,v,o}\fmfleft{v1}\fmfright{v2}\fmf{dashes}{v,v1}\fmf{dbl_plain,left}{v1,v2}\fmf{dashes}{v2,v}\fmf{boson}{i,v,o}\fmfdot{v1,v2,v}
\end{fmfgraph}&=-\frac{\ka^2}{4} {T_\mu}^\mu \frac{\de-2}{D-2} \sum_{\vec{n}} \pint{k} \frac{m^2_{\vec{n}}}{k^2(k^2-m^2_{\vec{n}})} \\
 \begin{fmfgraph}(30,30)
 \fmfbottom{i,v,o}\fmfleft{v1}\fmfright{v2}\fmf{dashes}{v,v1}\fmf{dbl_curly,left}{v1,v2}\fmf{dashes}{v2,v}\fmf{boson}{i,v,o}\fmfdot{v1,v2,v}
\end{fmfgraph}&=\frac{\ka^2}{2} {T_\mu}^\mu \frac{\de}{d} \sum_{\vec{n}} \pint{k} \frac{1}{k^2-m^2_{\vec{n}}}
\end{aligned}
\end{equation}
for the parachute shaped graphs and 
\begin{equation}\label{eq:new_moon}
\begin{aligned}
\begin{fmfgraph}(30,30)
 \fmfbottom{i,v,o}\fmftop{t}\fmf{dashes,left}{v,t}\fmf{dbl_wiggly,left}{t,v}\fmf{boson}{i,v,o}\fmfdot{t,v}
\end{fmfgraph}+
\begin{fmfgraph}(30,30)
 \fmfbottom{i,v,o}\fmftop{t}\fmf{dashes,left}{v,t}\fmf{dbl_plain,left}{t,v}\fmf{boson}{i,v,o}\fmfdot{t,v}
\end{fmfgraph}&=\frac{\ka^2}{4} {T_\mu}^\mu \frac{\de-2}{D-2} \sum_{\vec{n}} \pint{k} \frac{m^2_{\vec{n}}}{k^2(k^2-m^2_{\vec{n}})} \\
\begin{fmfgraph}(30,30)
 \fmfbottom{i,v,o}\fmftop{t}\fmf{dashes,left}{v,t}\fmf{dbl_curly,left}{t,v}\fmf{boson}{i,v,o}\fmfdot{t,v}
\end{fmfgraph}&=-\frac{\ka^2}{2} {T_\mu}^\mu \frac{\de}{d} \sum_{\vec{n}} \pint{k} \frac{1}{k^2-m^2_{\vec{n}}}
\end{aligned}
\end{equation}\fmfcurved
for the new~moon shaped graphs.  Here $T^{\mu}_{\mu}$ is the trace of the energy momentum tensor and the sum runs over the Kaluza-Klein~gravitons.
Clearly, the formal expressions for the momentum integrals in equations \eqref{eq:parachute} and \eqref{eq:new_moon}  are understood to be suitably regularized.
From the above expressions it directly follows that the sum of all branon tadpole graphs vanishes independently of the
size of the compactified dimensions. Most interestingly, to lowest order in the brane tension, there appear no
branon effects.

In the next step, let us consider the contribution of the Kaluza-Klein~gravitons to the gauge boson (gluon) polarization tensor.
Neglecting for a moment possible higher order derivative terms of the form $q^2(q^2\eta^{\mu\nu}-q^\mu q^\nu)$,
we obtain for the leading propagator divergence to order $\ka^2$:
\begin{multline}
\raisebox{-1.5ex}[1ex][0ex]{\begin{fmfgraph}(40,20)
\fmfleft{i}
\fmfright{o}
\fmf{boson}{i,v1}
\fmf{dbl_wiggly,left,tension=0.3}{v1,v2}
\fmf{boson,right,tension=0.3}{v1,v2}
\fmf{boson}{v2,o}
\fmfdot{v1,v2}
\end{fmfgraph}}+
\raisebox{-1.5ex}[1ex][0ex]{\begin{fmfgraph}(40,20)
\fmfleft{i}
\fmfright{o}
\fmf{boson}{i,v1}
\fmf{dbl_plain,left,tension=0.3}{v1,v2}
\fmf{boson,right,tension=0.3}{v1,v2}
\fmf{boson}{v2,o}
\fmfdot{v1,v2}
\end{fmfgraph}}=\ka^2\de^{ab}(q^2\eta^{\mu\nu}-q^\mu q^\nu)\times \\
	\times \frac{1}{2d}\left( 8-5d+\frac{(d-4)^2}{D-2}\right) \sum_{\vec{n}} \pint{k} \frac{1}{k^2-m^2_{\vec{n}}}\,.
\end{multline}
The last term in the bracket, $\sim1/(D-2)$ -- coming form the combination of the last term in the graviton propagator and the graviscalar -- is the only one proportional to the trace of the energy momentum tensor and thus vanishes in $d\!=\!4$ dimensions, where the Yang-Mills theory is classically conformal. The tadpole graphs yield
\begin{multline}
\frac{1}{2}
\begin{fmfgraph}(40,20)
\fmfbottom{i,o}
\fmftop{t}
\fmf{boson}{i,v1}
\fmf{boson}{v1,o}
\fmffreeze
\fmf{dbl_wiggly,left,tension=0.3}{v1,t,v1}
\fmfdot{v1}
\end{fmfgraph}
+\frac{1}{2}
 \begin{fmfgraph}(40,20)
\fmfbottom{i,o}
\fmftop{t}
\fmf{boson}{i,v1}
\fmf{boson}{v1,o}
\fmffreeze
\fmf{dbl_plain,left,tension=0.3}{v1,t,v1}
\fmfdot{v1}
\end{fmfgraph}=\ka^2\de^{ab}(q^2\eta^{\mu\nu}-q^\mu q^\nu) \times \\
	\times \frac{1}{8}\left( -d^2+8d-4-\frac{(d-4)(d-6)}{(D-2)}\right) \sum_{\vec{n}} \pint{k} \frac{1}{k^2-m^2_{\vec{n}}}\,.
\end{multline}
Again only the $1/(D-2)$ term vanishes in $d\!=\!4$ dimensions.
Finally their sum
\begin{multline}\label{eq:prop_corr}
\raisebox{.5ex}{\begin{fmfgraph*}(40,10)
\fmfbottom{i,o}
\fmftop{t}
\fmf{boson}{i,v,o}
\fmfv{label={$\scriptstyle \ka^2$},label.dist=0,decoration.shape=circle,decoration.filled=empty}{v}
\end{fmfgraph*}}=\ka^2\de^{ab}(q^2\eta^{\mu\nu}-q^\mu q^\nu)\times \\
	\times \frac{d-4}{8d} \left( -d^2+4d-8-\frac{(d-2)(d-8)}{(D-2)}\right)  \sum_{\vec{n}} \pint{k} \frac{1}{k^2-m^2_{\vec{n}}}
\end{multline}
manifests an overall factor of $(d\!-\!4)$ generalizing our zero result in \cite{EPR}, where only the pure 4-di\-men\-sion\-al case was considered. It is noteworthy that the vanishing of the gravitational correction in the $d\!=\!4$ case \emph{cannot} be explained by the tracelessness of the energy momentum tensor of gauge theories. As mentioned above, this argument holds only for the term $\sim1/(D-2)$ in \eqref{eq:prop_corr}.

The resulting divergence can be cancelled by the counter-term:
\begin{equation}
\raisebox{.5ex}{\begin{fmfgraph}(40,10)
\fmfbottom{i,o}
\fmftop{t}
\fmf{boson}{i,v,o}
\fmfv{decoration.shape=cross}{v}
\fmffreeze
\fmfiv{d.sh=circle,d.filled=empty}{vloc(__v)}
\end{fmfgraph}}=-i\de^{ab}(q^2\eta^{\mu\nu}-q^\mu q^\nu)\de_2,
\end{equation}
with $\de_2=Z_2-1$, $Z_2$ being the gluon wave function renormalization constant, and
\begin{equation}
\label{hier1}
 \de_2\Bigr|_{\O(\ka^2)}=\ka^2\frac{d-4}{8(D-2)}\left((d-3)(d-2)+\frac{\de}{d}(d^2-4d+8)\right) i\sum_{\vec{n}} \pint{k} \frac{1}{k^2-m^2_{\vec{n}}}\,.
\end{equation}

\begin{figure}[t]
\label{fig:grav_vert_corr}
\begin{center}
\begin{tabular}{c@{\quad}c@{\quad}c@{\quad}}
\begin{fmfgraph}(50,50)
\fmftop{t}
\fmfbottom{i,o}
\fmf{boson,tension=1.5}{t,v2}
\fmf{boson,tension=1.5}{i,v1}
\fmf{boson,tension=1.5}{v3,o}
\fmf{boson}{v1,v2}
\fmf{boson}{v2,v3}
\fmf{dbl_wiggly,tension=0}{v1,v3}
\fmfdot{v1,v2,v3}
\end{fmfgraph}&
\begin{fmfgraph}(50,50)
\fmftop{t}
\fmfbottom{i,o}
\fmf{boson}{t,v2}
\fmf{phantom}{i,v2}
\fmf{boson}{v2,o}
\fmffreeze
\fmf{boson,tension=1}{i,v1}
\fmf{boson,tension=0.4,right}{v1,v2}
\fmf{dbl_wiggly,tension=0.4,left}{v1,v2}
\fmfdot{v1,v2}
\end{fmfgraph}&
\raisebox{4ex}{$\dfrac{1}{2}$}
\begin{fmfgraph}(50,50)
\fmftop{t}
\fmfbottom{i,o}
\fmf{boson}{t,v}
\fmf{boson}{i,v,o}
\fmf{dbl_wiggly,left,tension=0.7}{v,v}
\fmfdot{v}
\end{fmfgraph}\\
\begin{fmfgraph}(50,50)
\fmftop{t}
\fmfbottom{i,o}
\fmf{boson,tension=1.5}{t,v2}
\fmf{boson,tension=1.5}{i,v1}
\fmf{boson,tension=1.5}{v3,o}
\fmf{boson}{v1,v2}
\fmf{boson}{v2,v3}
\fmf{dbl_plain,tension=0}{v1,v3}
\fmfdot{v1,v2,v3}
\end{fmfgraph}&
\begin{fmfgraph}(50,50)
\fmftop{t}
\fmfbottom{i,o}
\fmf{boson}{t,v2}
\fmf{phantom}{i,v2}
\fmf{boson}{v2,o}
\fmffreeze
\fmf{boson,tension=1}{i,v1}
\fmf{boson,tension=0.4,right}{v1,v2}
\fmf{dbl_plain,tension=0.4,left}{v1,v2}
\fmfdot{v1,v2}
\end{fmfgraph}&
\raisebox{4ex}{$\dfrac{1}{2}$}
\begin{fmfgraph}(50,50)
\fmftop{t}
\fmfbottom{i,o}
\fmf{boson}{t,v}
\fmf{boson}{i,v,o}
\fmf{dbl_plain,left,tension=0.7}{v,v}
\fmfdot{v}
\end{fmfgraph}
\end{tabular}
\end{center}
\caption{Gravitational  one-loop diagrams for the three-gluon function.\vspace{-2ex}}
\end{figure}
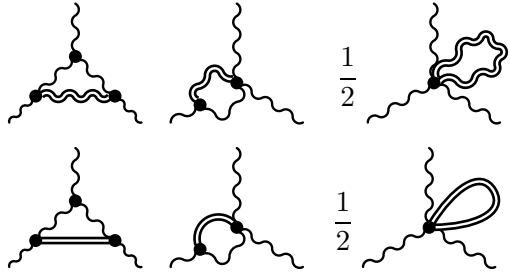
To determine the running of the coupling constant also the leading gravity induced divergencies of the 
amputated three-gluon function from figure~2 are needed. We find that these 
are cancelled by the three-gluon coun\-ter-term
\begin{equation}
\begin{aligned}
\raisebox{-6ex}[5ex][0ex]{\begin{fmfgraph}(50,50)
\fmftop{t}
\fmflabel{$\mu\:a$}{t}
\fmfbottom{l,r}
\fmflabel{$\nu\:b$}{l}
\fmflabel{$\rho\:c$}{r}
\fmf{boson,label=$p$}{t,v}
\fmf{boson,label=$q$}{l,v}
\fmf{boson,label=$k$}{v,r}
\fmfv{d.sh=cross,l=$g$,l.a=30,l.d=10}{v}
\fmffreeze
\fmfiv{d.sh=circle,d.filled=empty}{vloc(__v)}
\end{fmfgraph}}\hspace{-1em}=gf^{abc}\bigl[&\eta^{\mu\nu}(p-q)^\rho \\ +&\eta^{\nu\rho}(q-k)^\mu  \\ +&\eta^{\rho\mu}(k-p)^\nu \bigr] \de^{3g}_1 
\end{aligned}
\end{equation}
with the value
\begin{equation}
 \de_1^{3g}\Bigr|_{\O(\ka^2)}=\ka^2\frac{d-4}{8(D-2)}\left((d-3)(d-2)+\frac{\de}{d}(d^2-4d+8)\right) i\sum_{\vec{n}} \pint{k} \frac{1}{k^2-m^2_{\vec{n}}}\, .
\end{equation}
Note that this is identical to the gravitational contribution to the value of the two-point counter-term
constant $\de_2$ of (\ref{hier1}).
The purely gauge part of the one-loop divergencies are of course left unmodified.

The equality of the vertex and propagator correction is a direct consequence of gauge invariance\footnote{Here we thank Theodor Schuster for valuable comments.}. Due to the universality of the gauge coupling $g$,
its renormalization constants obtained from the three-gluon and gluon-ghost vertex must be the same, i.\,e.
\begin{equation}
\label{hier2}
 \frac{Z_1}{Z^{3/2}_2}=\frac{\tilde{Z}_{1}}{Z^{1/2}_2 \tilde{Z}_{2}}\,,
\end{equation}
where $Z_1(\tilde{Z}_1)$ denote the vertex renormalization constants for gluon (ghost) couplings.
Since the gluon ghosts are introduced after the expansion of the metric and thus do not couple to gravitons, we have
with $\tilde Z_1=1+\tilde\delta_1$ and $\tilde Z_2=1+\tilde\delta_2$ that 
\begin{align*}
 \tilde{\delta}_1\bigr|_{\O(\ka^2)} &= \tilde{\delta}_2\bigr|_{\O(\ka^2)} =0 
\intertext{and so by virtue of (\ref{hier2}) at one-loop level}
	 \delta^{3g}_1\Bigr|_{\O(\ka^2)} &=\delta_2\Bigr|_{\O(\ka^2)}\,.
\end{align*}

In order to extract the leading divergence from the loop-integrals one may 
replace the sum over discrete toroidal KK-modes $\vec{n}=(n_1,n_2,...,n_{\de})$ 
by an integration over the mass density
\begin{align*}
 i\sum_{\vec{n}} \pint{k} \frac{1}{k^2-m^2_{\vec{n}}} &= R^\de\int \diff[\de]{m} \pint{k} \frac{i}{k^2-m^2_{\vec{n}}} \\
                    &= R^\de \int \frac{\diff[D]{K}}{(2\pi)^d}\frac{i}{K^2} \\
                    &= \frac{2\pi^{D/2}}{(2\pi)^d\Ga(\frac{D}{2})}\frac{R^\de}{D-2}\left(\La^{D-2}-\mu^{D-2}\right),
\end{align*}
where $K=(k,m_{\vec{n}})$ is the $D$-dimensional momentum vector. Above we have also introduced the $D$ dimensional
UV-cut-off $\Lambda$ and a low energy reference scale $\mu$.
Taking further into account that
\begin{equation}
 V_{\de} \ka^2 =\kaD[2]=\frac{32\pi}{\emD[D-2]},
\end{equation}
we obtain
\begin{equation}
\begin{aligned}
 \de_1^{3g} \Bigr|_{\O(\ka^2)}&=\de_2\Bigr|_{\O(\ka^2)}=\Delta \frac{\La^{D-2}-\mu^{D-2}}{\emD[D-2]}, \\
 \Delta&=\frac{2}{(4\pi)^{D/2-1}\Ga(\frac{D}{2})}\frac{d-4}{(D-2)^2}\left((d-3)(d-2)+\frac{\de}{d}(d^2-4d+8)\right).
\end{aligned}
\end{equation}
The resulting gravitational contribution to the Yang-Mills coupling beta-function then reads
\begin{align}
 \be_g  &= g\frac{\del }{\del \log \mu}\left( \frac{3}{2}\de_2 - \de^{3g}_1 \right) \notag \\
 \Rightarrow\be_g \Bigr|_{\O(\ka^2)} &= -\frac{g}{(4\pi)^{D/2-1} \Ga(\frac{D}{2})}\frac{d-4}{D-2}\left((d-3)(d-2)+\frac{\de}{d}(d^2-4d+8)\right)\frac{\mu^{D-2}}{\emD[D-2]}
\end{align}

Irrespective of the presence of higher dimensional gravity, for $d\!=\!4$ (3-branes) there are 
no gravitational corrections to the Yang-Mills $\beta$-function  at one-loop order
\footnote{Note that for $d\!\neq\!4$ this result is specific for the de~Donder gauge ($\alpha\!=\!1$). Due to its non-zero mass dimension, the coupling constant and its renormalization might depend on the chosen gauge in the gravity sector, see e.\,g. \cite{ContRatt,Antoniadis:1985ub}.}.
It is crucial that the vanishing of the leading gravitational divergence and thus the absence of 
a gravitational induced running of the coupling constant is a unique feature of 4-dimensional gauge 
theories, independent of the number of extra dimensions. 

We claim that this is a gauge condition independent result based on the arguments of \cite{ContRatt} for the dimensionless gauge coupling in $d\!=\!4$ dimensions.

\subsection{Non-Abelian higher-dimensional counterterms}
In \cite{EPR} we studied in addition to the running of the Yang-Mills coupling constant the generation of gauge invariant terms whose mass dimension is six in $d\!=\!4$ dimensions. 
Generally, the one-loop gravitational UV-divergencies renormalize also gauge invariant terms of higher mass dimension which consequently should be included in the effective Lagrangian of the Yang-Mills sector 
\[ \label{eq:LYM_eff}
 \L_\text{eff.\,YM} = -\tfrac{1}{2}\tr\left[F^{\mu\nu}F_{\mu\nu}\right]  + a \tr\left[ D^\mu F_{\mu\rho} D_\nu F^{\nu\rho} \right]
        +b \tr\left[ {F^\al}_\be{F^\be}_\ga{F^\ga}_\al \right]\,,
\]
where $a,b$ are new effective coupling constants.
We found that in four dimensions only the coefficient $a$ is affected. This is an interesting result
because it is the non-abelian generalization of the
 Lee-Wick term \cite{LeeWick} recently considered in \cite{Grinstein} as part of an extented standard model\footnote{Contrary to the remarks in \cite{EPR}, \cite{Deser}, this term cannot
be simply rempoved by a field
redefinition, since such a non-linear field redefinition would introduce new ghost fields via the
Jacobian in the path integral.}.

The generalisation of our previous result to a
$(d\!+\!\de)$-dimensional braneworld hold a major difficulty. The
corresponding gravitational contributions are no longer logarithmic
but powerlike divergent in $D\!\neq\!4$ and non-leading powerlike
divergencies depend on the chosen parametrization of the loop momentum
if cut-off regularization is applied. Figure~3
shows a general parameterization for a self-energy
diagram. Here $k$ is loop-momentum and $x\!\in\!\mathbb{R}$ is the
fraction of the outer momentum $q$ flowing on the graviton
line. As explict
calculations show the results for the sub-leading, non-logarithmical
divergent contributions depend on $x$. More complex graphs, like triangles etc., have to be parametrized by
more such free parameters, one for each independent external momentum.
\begin{figure}[t]
\begin{center}
\raisebox{-0.6cm}{
\includegraphics[height=4.0cm]{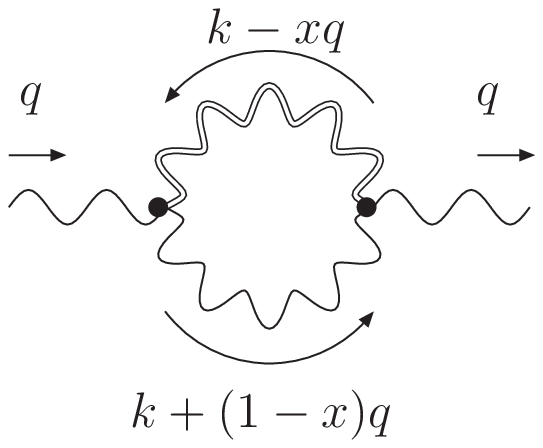}}
\end{center}
\caption{A general parameterization of the loop-momenta for a bubble graph.}
\label{fig:bubble_mom}
\end{figure}

This obstacle can be partially overcome by demanding gauge invariant
results which fixes some of the degrees of freedom. The remaining ambiguity can
be eliminated by requiring a universally applicable parameterization,
i.\,e. \emph{all} bubbles, triangles etc. parameterized in identical
manner. The consideration of similar problems involving fermions and
charged scalars in \cite{RodSchu} shows that gauge invariant
counter-terms are achieved only by parameterizations, where the
graviton propagator does not carry any part of the external momenta,
i.\,e. the graviton momentum is exactly the loop-momentum which one is
integrated over. Using the above notation this means e.\,g. $x=0$ for
the bubble graphs.

Applying this prescription, our calculations led to the following expressions of the corresponding counterterms
\[
 \L_\text{c.\,t.} = \de_{a} \tr\left[ D^\mu F_{\mu\rho} D_\nu F^{\nu\rho} \right]
        +\de_{b} \tr\left[ {F^\al}_\be{F^\be}_\ga{F^\ga}_\al \right]
\]
with
\begin{align}
 \de_{a} &= \ka^2\frac{(d-2)^2}{2(d+2)(D-2)}\left(3-d+\frac{\de}{d}(4-\tfrac{3}{2}d)\right) i\sum_{\vec{n}} \pint{k} \frac{1}{k^2(k^2-m^2_{\vec{n}})}\,, \\
 \de_{b} &= g\ka^2\frac{(d-4)}{d(d+2)(D-2)}\left(8-8d-d^2-\de(16+d)\right) i\sum_{\vec{n}} \pint{k} \frac{1}{k^2(k^2-m^2_{\vec{n}})}
\end{align}
or equivalently, by calculating again the sum and integrals
and introducing new dimensionless tilded quantities $a=\tilde{a}/\emD[2]$ and $b=\tilde{b}\,g/\emD[2]$ and the coresponding counter-terms $\de_a=\de_{\tilde{a}} /\emD[2]$ and $\de_b=\de_{\tilde{b}} g/\emD[2]$,
\begin{align}
 \de_{\tilde{a}} &=\frac{8}{(4\pi)^{D/2-1}\Ga(\tfrac{D}{2})}\frac{(d-2)}{(d+2)(D-4)}\left(3-d+\frac{\de}{d}(4-\tfrac{3}{2}d)\right)\frac{\La^{D-4}-\mu^{D-4}}{\emD[D-4]} \,, \\
 \de_{\tilde{b}} &=\frac{16}{(4\pi)^{D/2-1}\Ga(\tfrac{D}{2})}\frac{(d-4)\left(8-8d-d^2-\de(16+d)\right)}{d(d+2)(d-2)(D-4)}\frac{\La^{D-4}-\mu^{D-4}}{\emD[D-4]} \,,
\end{align}

Note that the counterterm $\de_{b}$ vanishes for $d\!=\!4$ in agreement with our previous result\cite{EPR}. Thus, there is no need to
consider a term $\sim \tr FFF$ in usual four space-time dimensions.
This is a welcomed result, since such a term would lead to a violation of unitarity \cite{Grinstein}.

Most interestingly, due to higher-dimensional gravity, for $d=4$ the nonvanishing counterterm $\de_{a}$
depends now on the fundamental gravitational constant $\ka_{(D)}$ which is assumed to be much larger than
$\ka$.

\newpage
\section{Summary and Conclusions}
In this work we have applied the techniques of effective field theory in the spirit of \cite{donoghue} 
for investigating the higher dimensional Einstein-Yang-Mills system in a $D=d +\delta$ large extra dimensional
brane world. In this scenario the gauge bosons live on a
fluctuating $(d\!-\!1)$-brane and gravitons move freely in the compactified
$\delta$ extra dimensions.
Following standard methods \cite{ContRatt,GiuRattWells,HanLykkZh} by
expanding the $D$-dimensional metric around a flat space-time background
in the graviton field $h_{MN}$,
then performing the Kaluza-Klein (KK) reduction and adding the gauge boson part on the brane, 
we obtained the neccessary Feynman rules for propagators
and interaction vertices of gauge bosons, KK-gravitons and branons.
On this basis we then performed one-loop calculations including towers of excited KK-gravitons in 
order to determine the possible gravity-induced power-law corrections to the Yang-Mills $\beta$-function and 
the generation of higher derivative operators. As we found, all tadpole contributions to the gauge boson
propagator, which are induced by branon-KK-graviton mixing, vanish in the considered lowest order expansion
in the brane tension. Again, as in our earlier paper \cite{EPR}, for the physical 3-brane (d=4) the possible power-law
behaviour of the running gauge coupling cancels. However, as an important new effect of the incorporation of the
ADD-scenario of large size (compact) extra dimensions into the Einstein-Yang-Mills system,
we now obtain a significant increase of the gravity-induced structure coefficient $a$ of the non-abelian 
counterterm  $a \sim \tr (DF)^2$. Clearly, this is a direct consequence of the
lowering of the D-dimensional Planck scale $\emD$. Interestingly, the gravitationally induced term is of the non-abelian
Lee-Wick form, which has been discussed recently as a mechanism to stabilize the Higgs mass \cite{Grinstein}. 
Irrespective of this, the lowered mass scale provides a window for possible to observation of the 
associated massive Lee-Wick vector-bosons at TeV energies -- iff large extra dimensions exist.

The methods of effective field theory used in this paper can only be applied for computing contributions to physical
processes at low-energy scales of external momenta. When applying effective field theory to large virtual momenta in
the considered gravitational one-loop calculations, it cannot really be controlled. Thus, one has to introduce an explicit
UV-cutoff  $\Lambda$ which has to be treated as a phenomenological parameter \cite{ContRatt}. In this sense
the main goal of the present work was indeed a more conceptual one by studying the effect of large extra dimensions on
new gravity-induced structures like a possible power-law running of the gauge boson coupling and a qualitative
estimate of the structure constant in the induced counterterms. The techniques developed in this paper may
turn out to be useful in future investigations.

\acknowledgments
We thank D.\,I.~Kazakov, J.~Pawlowski, A.\,A. Slavnov, T. Schuster, D.\,V. Shirkov and  O.\,E. Teryaev for 
discussions and critical remarks. One of us (D.\,E.) is grateful to V.\,V. Voronov and the other 
colleagues of the Bogoliubov Laboratory for Theoretical Physics at JINR Dubna for kind hospitality
and to the Bundesministerium f\"ur Bildung und Forschung for financial support. This work was supported by the
Volkswagen Foundation.
Our computation made use of the symbolic manipulation system Form \cite{Form}.
\appendix

\section{Validity of the Energy Expansion}\label{appendix}
In our investigation of possible one-loop contributions from gravitons and branons we focused only on gravity effects $\sim \kappa^2$ and not on effects of the brane tension $\tau$, i.\,e. neglecting both terms $\O(\tau)$ and $\O(\tau^{-1})$. In this section we show that an energy hierarchy allowing this ansatz exists, due to limited number of combinations of energy scales appearing in the expansion.
The various energy scales of the effective field theory are:
\begin{equation*}
 \kaD \simeq \emD[\frac{2-D}{2}]\,, \qquad \tau = M_\tau^d\,, \qquad E\,,\qquad M_V=(2\pi R)^{-1} .
\end{equation*}
For the effective theory of gravity to be valid, we need the following requirements:
\begin{enumerate}
 \item Of course, the expansion parameter itself must be small. Keeping the number of incoming and outgoing particles, fixed the expansion parameter is
    \begin{equation}
     \kaD E^\frac{D-2}{2} = \left(\frac{E}{\emD}\right)^\frac{D-2}{2} \ll 1   \quad\Longrightarrow\quad    \frac{E}{\emD} \ll 1 \,. \label{eq:ellmp}
    \end{equation}
 \item  In our consideration we neglected higher branon corrections, e.\,g.\begin{fmfgraph}(30,15)\fmfbottom{i,o}\fmftop{t}\fmf{dashes}{i,v,o}\fmffreeze\fmf{dashes,left}{v,t,v}\fmfdot{v}\end{fmfgraph}, which are at least $\O(\tau^{-1})$.
	Thus we need
    \begin{equation}
     \frac{E^d}{\tau} = \left(\frac{E}{M_\tau}\right)^d \ll 1 \quad\Longrightarrow\quad    \frac{E}{M_\tau} \ll 1 \,. \label{eq:qllmt}
    \end{equation}
 \item  As a consequence of \eqref{eq:brane_tension_exp} the effective theory includes tree-level mixing between Kaluza-Klein graviton states, direct
 \begin{fmfgraph}(30,10)\fmfleft{i}\fmfright{o}\fmf{dbl_wiggly}{i,v,o}\fmfv{d.sh=circle,d.f=empty,d.size=5}{v}\end{fmfgraph} and indirect via branons
 \begin{fmfgraph}(40,10)\fmfleft{i}\fmfright{o}\fmf{dbl_wiggly}{i,v1}\fmf{dashes,dash_len=1}{v1,v2}\fmf{dbl_wiggly}{v2,o}\fmfvn{d.sh=circle,d.f=empty,d.size=5}{v}{2}\end{fmfgraph}, 
 as well as tadpoles \raisebox{-10\unitlength}[0ex][0ex]{\begin{fmfgraph}(30,20)\fmfleft{i}\fmfright{o}\fmftop{t}\fmf{boson}{i,v,o}\fmfdot{v}\fmffreeze\fmf{dbl_wiggly}{v,t}\fmfv{d.sh=circle,d.f=empty,d.size=5}{t}\end{fmfgraph}}.
 All of these effects are of order $\O(\kaD[2]\tau)$. Neglecting their effect requires
    \begin{equation}
     \kaD[2] \tau = \frac{M_\tau^d}{\emD[D-2]} = \emD[2] \frac{M_\tau^d}{\emD[D]} \ll E^{2-D+d}=E^{2-\de} \,. \label{eq:mtcond}
    \end{equation}
 To find a consistent energy hierarchy in which the mixing can be ignored, we have to consider the following cases:
 \begin{enumerate}
     \item $\de>2$, assuming $M_\tau \gg \emD$ \eqref{eq:mtcond} is equvialent to:
         \begin{align*}
          \quad E^{\de-2} \ll \frac{\emD[D-2]}{M_\tau^d} \stackrel{M_{(\!D\!)} \ll M_\tau}{\ll} M_\tau^{D-2-d} =M_\tau^{\de-2} ,
 	 \end{align*}
	i.\,e. $M_\tau \gg E$, in agreement with the requirement \eqref{eq:mtcond}.\\
        The assumption $M_\tau \gg \emD$ is feasible, since only the brane density $M_\tau^d \rho^{-\de}$, with $\rho$ being the thickness, is bounded by the $D$ dimensional Planck scale $\emD[D]$
     \item In the case $\de=2$ \eqref{eq:mtcond} simplifies to
	\[
	 M_\tau \ll \emD \,.
	\]
     \item Finally, if $\de=1$ \eqref{eq:mtcond} is equvialent to
        \[
         M_\tau^d \ll E\emD[d-1] \ll \emD
        \]
        but it should still be $E^1\emD[d-1] \gg E^d$, so that \eqref{eq:qllmt} can be satisfied.
    \end{enumerate}
\end{enumerate}
Thus we find the energy hierarchy:
\begin{align}
 \text{for~}\de>2   && E &\ll \emD \ll M_\tau  \,, \\
 \text{for~}\de=2    && E &\ll M_\tau \ll \emD  \\
 \text{and for~}\de=1    && E &\ll M_\tau \ll E^{1/d}\emD[{(d-1)/d}] \ll \emD
\end{align}
in which our pertubative considerations above are meaningful.

\end{fmffile}

\begin{thebibliography}{99}
\bibitem{ADD} N.~Arkani-Hamed, S.~Dimopoulos and G.~Dvali, 
  ``The hierarchy problem and new dimensions at a millimeter,''
  \plb{429}{1998}{263}, \hepph{9803315}.
\bibitem{Reviews}  V.~A.~Rubakov,
  ``Large and infinite extra dimensions: An introduction,''
  \textit{Phys.\ Usp.\  }{\bf 44} (2001) 871,
  \ufn{171}{2001}{913},
  \hepph{0104152}; \\
G.~Gabadadze,
  ``ICTP lectures on large extra dimensions,''
  \hepph{0308112}; \\
R.~Sundrum,
  ``To the fifth dimension and back. (TASI 2004),''
  \hepth{0508134}; \\
 R.~Rattazzi,
  ``Cargese lectures on extra dimensions,''
  \hepph{0607055}.
\bibitem{PDG} [Particle Data Group],
C. Amsler et al., \plb{667}{2008}{1}. 
\bibitem{AADD} I.~Antoniadis,  N.~Arkani-Hamed, S.~Dimopoulos and G.~Dvali, 
  ``New dimensions at a millimeter to a Fermi and superstrings at a TeV,''
  \plb{436}{1998}{257}, \hepph{9804398}.
\bibitem{Antoniadis:1990ew}
  I.~Antoniadis,
  ``A Possible new dimension at a few TeV,''
  \plb{246}{1990}{377}.
\bibitem{Dienes}K.~R.~Dienes, E.~Dudas and T.~Gherghetta,
  ``Grand unification at intermediate mass scales through extra dimensions,''
  \npb{537}{1999}{47},
  \hepph{9806292}.
\bibitem{RobWil} S.P.~Robinson, F.~Wilczek, 
  ``Gravitational correction to running of gauge couplings,''
  \prl{96}{2006}{231601}, \hepth{0509050}.
\bibitem{Pietrykowski} A.R.~Pietrykowski, 
  ``Gauge dependence of gravitational correction to running of gauge couplings,''
  \prl{98}{2007}{061801}, \hepth{0606208}.
\bibitem{Toms} D.J.~Toms,  
  ``Quantum gravity and charge renormalization,''
  \prd{76}{2007}{045015},
  \arXivid{0708.2990}~{\tt [hep-th]}.
\bibitem{EPR} D.~Ebert, J.~Plefka, A.~Rodigast, 
  ``Absence of gravitational contributions to the running Yang-Mills coupling,''
  \plb{660}{2008}{579}, \arXivid{0710.1002}~{\tt [hep-th]}.
\bibitem{Elias} E.~Kiritsis and C.~Kounnas,
  ``Infrared Regularization of Superstring Theory And The One Loop Calculation
  of Coupling Constants,''
  \npb{442}{1995}{472},
  \hepth{9501020}.
\bibitem{Deser}  S.~Deser, H.~S.~Tsao and P.~van Nieuwenhuizen,
  ``One Loop Divergences Of The Einstein Yang-Mills System,''
  \prd{10}{1974}{3337}.
\bibitem{LeeWick} T.~D.~Lee and G.~C.~Wick, 
  ``Negative Metric and the Unitarity of the S Matrix,''
  \npb{9}{1969}{209}; 
  ``Finite Theory of Quantum Electrodynamics,''
  \prd{2}{1970}{1033}.
\bibitem{Grinstein} B.~Grinstein, D.~O'Connell and M.~B.~Wise,
  ``The Lee-Wick standard model,''
  \prd{77}{2008}{025012}
  \arXivid{0704.1845}~{\tt [hep-ph]}.
\bibitem{WuZhong}
  F.~Wu and M.~Zhong,
  ``The Lee-Wick Fields out of Gravity,''
  \plb{659}{2008}{694}
  \arXivid{0705.3287}~{\tt [hep-ph]}.
\bibitem{ContRatt} R.~Contino, L.~Pilo, R.~Rattazzi, A.~Strumina, 
  ``Graviton loops and brane observables,''
  \jhep{06}{2001}{005}, \hepph{0103104}.
\bibitem{Antoniadis:1985ub}
  I.~Antoniadis, J.~Iliopoulos and T.~N.~Tomaras,
  ``Gauge Invariance In Quantum Gravity,''
  \npb{267}{1986}{497}.
\bibitem{GiuRattWells} G.\,F.~Guidice, R.~Rattazzi, J.\,D.~Wells,
  ``Quantum gravity and extra dimensions at high-energy colliders,''
   \npb{544}{1999}{3}, \hepph{9811291}.
\bibitem{HanLykkZh} T.~Han, J,\,D.~Lykken, R.-J.~Zhang,
  ``On Kaluza-Klein states from large extra dimensions,''
   \prd{59}{1999}{105006}, \hepph{9811350}.
\bibitem{Sundrum} R.~Sundrum,
  ``Effective field theory for a three-brane universe,''
   \prd{59}{1999}{085009}, \hepph{9805471}.
\bibitem{RodSchu} A.~Rodigast and T.~Schuster,
  ``No Lee-Wick Fields out of Gravity,''
  \prd{79}{2009}{125017}
  \arXivid{0903.3851}~{\tt [hep-ph]}.
\bibitem{donoghue} F.~Donoghue, ``General Relativity As An Effective Field Theory: The Leading Quantum
  Corrections,''\prd{50}{1994}{3874}.
\bibitem{Form} J.A.M.Vermaseren, ``New features of FORM,''\href{http://arxiv.org/abs/math-ph/0010025}{\tt math-ph/0010025}.
\end{thebibliography}
\end{document}